\documentclass[%
 reprint,
 amsmath,amssymb,
 aps,
]{revtex4-2}

\usepackage{graphicx}
\usepackage{dcolumn}
\usepackage{bm}
\usepackage{hyperref}
\hypersetup{colorlinks, linkcolor={blue}, citecolor={blue}, urlcolor={blue}}

\begin{document}
\title{Magnon-Driven Magnetothermal Transport in Magnetic Multilayers}

\author{Ping Tang$^1$}
\email{tang.ping.a2@tohoku.ac.jp}
\author{Ken-ichi Uchida$^{2,3}$}
\author{Gerrit E. W. Bauer$^{1,4,5}$}
\affiliation{$^1$WPI-AIMR, Tohoku University, Sendai 980-8577, Japan}
\affiliation{$^2$National Institute for Materials Science (NIMS), Tsukuba 305-0047, Japan}
\affiliation{$^3$Department of Advanced Materials Science, Graduate School of Frontier Sciences, The University of Tokyo, Kashiwa 277-8561, Japan}
\affiliation{$^4$Institute for Materials Research and CSIS, Tohoku University, Sendai 980-8577, Japan} 
\affiliation{$^5$Kavli Institute for Theoretical Sciences, University of the Chinese Academy of Sciences,
Beijing 10090, China}
\date{\today}

\begin{abstract}
All-solid-state nanoscale devices capable of efficiently controlling a heat flow are crucial for advanced thermal management technologies. Here we predict a magnon-driven magnetothermal resistance (mMTR) effect in multilayers of ferromagnets and normal metals, i.e. a thermal resistance that varies when switching between parallel and antiparallel magnetization orientations of the ferromagnetic layers, even in the absence of conduction electrons in the ferromagnets. The mMTR arises from an interfacial temperature drop caused by magnon spin accumulations and can be engineered by the layer thicknesses, spin diffusion lengths, and spin conductances. The mMTR predicted here enables magnetothermal switching in insulator-based systems; we already predict large mMTR ratios up to 40$\%$ for superlattices of the electrically insulating magnet yttrium iron garnet and elemental metals.
\end{abstract}
\maketitle

\emph{Introduction.---}The magneto-thermal resistance (MTR) effect enables the magnetic control of thermal transport. Solid-state devices with a high MTR ratio and a broad operating temperature range around room temperature are essential for thermal management technologies \cite{moore2014emerging}. A traditional approach focuses on the MTR of single-phase bulk materials, but it often requires impractically high magnetic fields or extremely low operation temperatures \cite{PhysRevB.55.15471,zhang2007giant,PhysRevB.77.054436,wang2010large,fu2020largely,hirata2023magneto,koster2024giant}. More promising is the giant MTR in ferromagnetic (FM)$\vert$nonmagnetic (NM) multilayers \cite{sato1993giant,sato1993huge, PhysRevB.54.15273,yang2004thermal,jeong2012giant, PhysRevB.87.134406,kimling2015spin,asam2019thermal,nakayama2021above}, the thermal version of the giant magnetoresistance (GMR) \cite{PhysRevLett.61.2472,binasch1989enhanced}, where the thermal conductivity varies with the magnetization configurations of the ferromagnetic layers, opening a pathway towards nonvolatile, nanoscale thermal switching devices \cite{arima2024observation}. Similar to the essential role of electronic spin accumulation in the GMR, spin-dependent heat accumulation or temperature of electrons is proposed to explain the MTR in magnetic multilayers \cite{dejene2013spin,kimling2015spin}. Because of the Wiedemann-Franz law and short spin-dependent heat relaxation length (compared to the spin diffusion length), the MTR has long been considered a byproduct of the GMR and has only received little attention. Surprisingly, several experimental groups report an MTR exceeding the GMR \cite{yang2004thermal,jeong2012giant,asam2019thermal,nakayama2021above}, implying an unconventional magneto-thermal transport mechanism.
\begin{figure}
\centering
\par
\includegraphics[width=8.6cm]{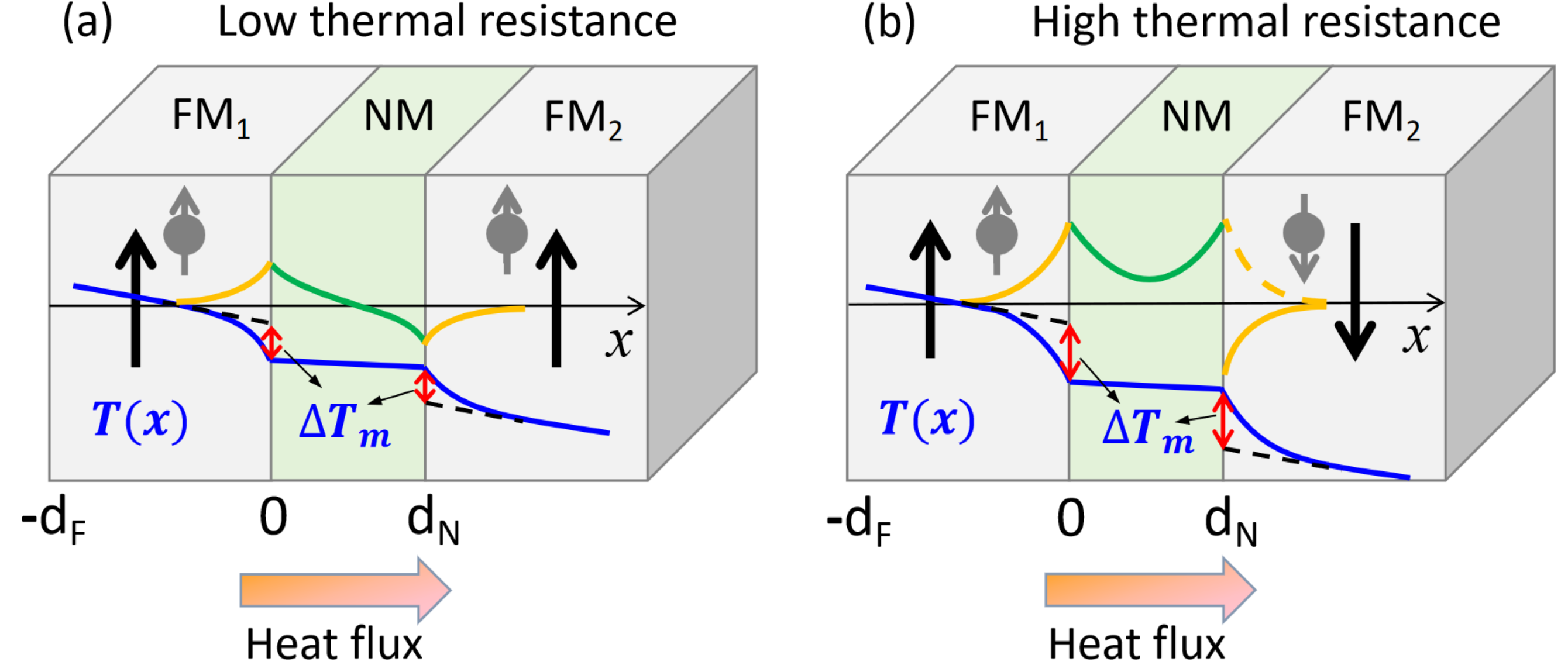}
\newline \caption{Concept of the magnon-driven thermal resistance in (FM$_1$)$\vert$nonmagnetic metal (NM)$\vert$ferromagnet (FM$_2$). A thermal bias excites non-equilibrium magnon accumulations (orange solid curves). Its gradient generates a back-flow heat current by diffusion and thereby, for a fixed heat current bias,  an additional temperature drop $\Delta T_{m}$  as sketched by the blue lines. (a) In the parallel configuration, the spin accumulations injected by the two FMs into the metal spacer cancel, corresponding to a high thermal conductance. In the antiparallel configuration (b), they add up to form a high-resistance state. The arrows on the grey dots represent the magnon spin angular momenta, the green solid curves in the NM layer the spin accumulation of electrons, and the dashed orange curves in (b) the magnonic spin accumulation opposite to the magnon number accumulation (solid orange curves).}%
\label{Fig-1}%
\end{figure}

Magnons, the collective quasi-particle excitations of magnetic order, serve as an additional carrier of energy and spin in a magnetic metal \cite{an2013unidirectional}. Under the gradient of a temperature or non-equilibrium magnon accumulation \cite{zhang2012magnon,rezende2014magnon,cornelissen2016magnon}, magnon diffusion contributes to both heat and spin currents, leading to magnetic-field dependent thermal conduction \cite{yelon1972magnon,hsu1976magnon, Rz2015} and the spin Seebeck effect \cite{uchida2010spin}, respectively. Recently, Hirai \emph{et al.} \cite{hirai2024} reported surprisingly large magnon contributions to the thermal transport properties of a ferromagnetic metal even at room temperature. 

Motivated by these experiments \cite{nakayama2021above,hirai2024} we present in this Letter a pronounced magnon-driven MTR (mMTR) in magnetic multilayers (see Fig.~\ref{Fig-1}): a thermally driven magnon current within the FM layers produces non-equilibrium magnon accumulation near FM$\vert$NM interfaces that depends on the interlayer magnetic configuration and causes an additional temperature drop due to a feedback diffusion heat current, giving rise to the MTR effect. This is analogous to the GMR with current perpendicular to the interfaces, in which interfacial spin accumulation of electrons plays an essential role \cite{gijs1997perpendicular}. We explain the large mMTR at room temperature by two factors. While in the GMR only electrons near the Fermi level play a role, the boson statistics of magnons allow all of them to contribute to transport. Moreover, the magnon diffusion length can exceed that of the mobile electrons by orders of magnitude. Since magnons can exist in magnetic insulators, the mMTR goes beyond the GMR mechanism, offering the potential for insulator-based magnetothermal switching devices.

\emph{Model.---}We address the mMTR for a spin valve of two identical FM layers and a thin NM spacer, as illustrated in Fig.~\ref{Fig-1}, as well as FM\(|\)NM multilayers. In a metallic FM, both electrons and magnons contribute to the MTR. The electron-induced MTR is caused by spin-dependent thermal conductivities in the FM and FM\(|\)NM interfaces and thereby directly related to the GMR by the Wiedemann-Franz law. Here we focus on electrically insulting FM layers to isolate a previously unrecognized magnonic mechanism. We disregard the coherent magnon coupling between FM layers by a non-local exchange interaction through ultrathin metal spacers that play a role only at low temperatures.    

In linear response a temperature (chemical potential) gradient $\partial T_{i}$ ($\partial\mu_{i}$) drives spin [$\mathcal{J}_{i}$] and heat [$Q_{i}$] currents in the $i$the FM layer \cite{cornelissen2016magnon}
\begin{equation}
\left(\begin{matrix}
  \frac{e}{\hbar}\mathcal{J}_{i}\\
 Q_{i}
\end{matrix}\right)=-\left(\begin{matrix}
\sigma_{m} & \mathcal{L}_{m}\\
T_{0}\mathcal{L}_{m} & \kappa_{F}
\end{matrix}\right) \left(\begin{matrix}
\partial\mu_{i}/e\\
\partial T_{i}
\end{matrix}\right), \label{FM}
\end{equation}
where $\sigma_{m}$ and $\mathcal{L}_{m}$ are the magnon spin conductivity and spin Seebeck coefficient, respectively, while $\kappa_{F}$ is the total thermal conductivity including phonon contributions. We employed the Onsager reciprocity of the off-diagonal magnon Seebeck and Peltier coefficients at the average temperature $T_{0}$. The polarization of the magnon spin current in Eq.~(\ref{FM}) is collinear with the magnetization direction of the $i$th FM layer and changes sign when reversing the magnetization. The spin [$\mathcal{J}_{N}$] and heat [$Q_{N}$] currents in the NM layer, on the other hand, obey
\begin{equation}
\left(\begin{matrix}
  \mathcal{J}_{N}\\
 Q_{N}
\end{matrix}\right)=-\left(\begin{matrix}
\frac{\hbar}{4e^{2}}\sigma_{N}& 0\\
0 & \kappa_{N}
\end{matrix}\right) \left(\begin{matrix}
\partial \mu_{N}\\
\partial T_{N}
\end{matrix}\right) \label{NM}
\end{equation}
where $\sigma_{N}$ and $\kappa_{N}$ are the electric and thermal conductivities of the metal, respectively, and $\mu_{N}$ the spin chemical potential, \textit{i.e.} the chemical potential difference of spin-up and -down electrons. In contrast to the FM layers, there are no off-diagonal elements.

Biasing the spin valve by a given heat current \(Q\) induces temperature and chemical potential gradients. Inverting the above relations
\begin{align}
\partial T_{i}(x)=&-\frac{Q}{\kappa_{F}}-\frac{T_{0}\mathcal{L}_{m}}{e\kappa_{F}}\partial\mu_{i}(x) \label{Ti}\\
\partial T_{N}(x)=&-\frac{Q}{\kappa_{N}},
\end{align}   
where magnon Peltier effect generates a non-equilibrium magnon accumulation and the second term in Eq.~(\ref{Ti}). The thermal resistivity \(\rho\) of the spin valve with spacer thickness \(d_{N}\) and equal thickness \(d_{F}\) of the two magnets reads
\begin{align}
\rho\equiv\frac{\Delta T}{LQ}=\frac{1}{L}\left(\frac{2d_{F}}{\kappa_{F}}+\frac{d_{N}}{\kappa_{N}}\right)+\rho_{m} \label{rho},
\end{align}
where $L=2d_{F}+d_{N}$, $\Delta T$ is the total temperature drop. Here we introduced a magnon-driven magnetothermal resistivity $\rho_{m}=\Delta T_{m}/(LQ)$, where
\begin{align}
\Delta T_{m}=\frac{T_{0}\mathcal{L}_{m}}{e\kappa_{F}} \int dx^{\prime} [\partial\mu_{1}(x^{\prime})+\partial\mu_{2}(x^{\prime})] \label{RM}
\end{align}
is the temperature drop in the presence of a magnon accumulation. In the following, we calculate Eq.~(\ref{RM}) and $\rho_{m}$ by spin diffusion theory when the two magnetizations are parallel and antiparallel.

\emph{Results.---}The solutions to the magnon diffusion equation \((\partial^2-\lambda_m^{-2})\mu_i=0\) are \cite{zhang2012magnon,rezende2014magnon,cornelissen2016magnon}, 
\begin{equation}
\mu_{i}(x)=A_{i}e^{x/\lambda_{m}}+B_{i}e^{-x/\lambda_{m}} ,\label{uF}
\end{equation}
where $A_{i}$ and $B_{i}$ are coefficients determined by boundary conditions and $\lambda_{m}$ the magnon diffusion length. The electronic spin chemical potential in the NM layer obeys
\begin{equation}
\mu_{N}(x)=C e^{x/\lambda_{N}}+De^{-x/\lambda_{N}} 
\end{equation}
where $\lambda_{N}$ is the spin diffusion length of electrons. In a collinear configuration without interfacial spin torques, the spin currents  at two NM-FM interfaces are conserved, \textit{i.e. }
\begin{align}
\mathcal{J}_{1}(0)=\mathcal{J}_{N}(0),\,\,\,\mathcal{J}_{N}(d_{N})=s\mathcal{J}_{2}(d_{N})
\end{align}
where $s=\pm$ accounts for the polarization of the magnonic spin current in the second FM layer, i.e., ($+$) when the magnetization of the second FM layer is aligned parallel  ($-$) when antiparallel to that of the first FM layer. Assuming transparent NM$|$FM interfaces \cite{Notecon}, the magnon and electron spin chemical potentials are also continuous at the interfaces with
\begin{align}
\mu_{1}(0)=\mu_{N}(0), \,\, \,\mu_{N}(d_{N})=s \mu_{2}(d_{N})
\end{align}
where $s=-1$ reflects the annihilation of magnons in the second FM by a spin-flip in the normal metal when its magnetization is antiparallel, i.e., ``magnon holes" with opposite spins are created in the second FM layer. When the FM thickness is smaller than or comparable to the magnon diffusion length we must specify the boundary conditions at the outer interfaces. In the following, we consider several experimentally relevant cases. 

\emph{(i) Spin-sinking contact.} We first consider a spin valve sandwiched between good spin sinks such as Pt, which implies that $\mu_{1}(-d_{F})=\mu_{2}(d_{N}+d_{F})=0$. The solution of Eq.~(\ref{uF}) with these boundary conditions leads to thermal resistivities
\begin{widetext}
  \begin{align}
 \rho_{m}^{\text{P}}=\frac{2\hbar}{e^2L}\frac{T_{0}\mathcal{L}_{m}^{2}}{ \kappa_{F}^2} \frac{G_{m}+G_{N}\tanh\frac{d_{F}}{\lambda_{m}}\tanh\frac{d_{N}}{2\lambda_{N}}}{\coth\frac{d_{F}}{\lambda_{m}}G_{m}^{2}+\tanh\frac{d_{F}}{\lambda_{m}}G_{N}^{2}+2G_{m}G_{N}\coth\frac{d_{N}}{\lambda_{N}}} \\ 
 \rho_{m}^{\text{AP}}=\frac{2\hbar}{e^2L}\frac{T_{0}\mathcal{L}_{m}^{2}}{ \kappa_{F}^2} \frac{G_{m}+G_{N}\tanh\frac{d_{F}}{\lambda_{m}}\coth\frac{d_{N}}{2\lambda_{N}}}{\coth\frac{d_{F}}{\lambda_{m}}G_{m}^{2}+\tanh\frac{d_{F}}{\lambda_{m}}G_{N}^{2}+2G_{m}G_{N}\coth\frac{d_{N}}{\lambda_{N}}}
\end{align}  
\end{widetext}
 for the parallel and antiparallel configurations, respectively. Here $G_{m}\equiv(\hbar/e^{2})\bar{\sigma}_{m}/\lambda_{m}$ and $G_{N}\equiv(\hbar/4e^{2})\sigma_{N}/\lambda_{N}$ are the spin conductances (per unit area) in the FM and NM, respectively, and
 \begin{align}
\bar{\sigma}_{m}\equiv\sigma_{m}\left(1-\frac{T_{0}\mathcal{L}_{m}^{2}}{\kappa_{F}\sigma_{m}} \right)=\sigma_{m}\left(1-zT_{m}\frac{\kappa_{m}}{\kappa_{F}} \right)
 \end{align}
is the magnon spin conductivity corrected for the spin Seebeck/Peltier effects, where $zT_{m}\equiv T_0\mathcal{L}_{m}^{2}/(\kappa_{m}\sigma_{m})$ is a thermomagnonic figure of merit and $\kappa_{m}$ the magnon thermal conductivity. Here $\rho_{m}^{\text{P}}$ and $\rho_{m}^{\text{AP}}$ scale as $\kappa_{F}^{-2}$, implying that they can be enhanced by choosing FM layers with low thermal conductivities. 

\begin{figure}
\centering
\par
\includegraphics[width=8.6cm]{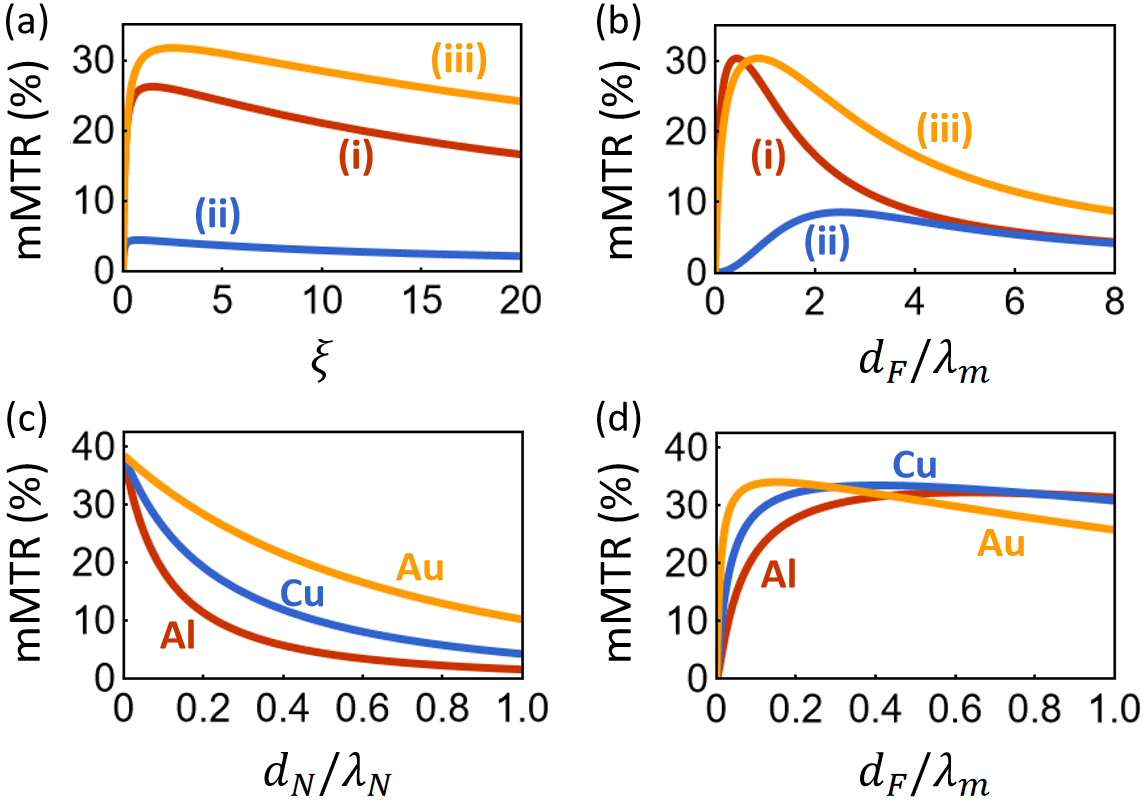}
\newline \caption{(a) and (b): The mMTR ratio as a function of $\xi$ and $d_{F}/\lambda_{m}$ of a YIG$|$NM$|$YIG  spin valve calculated for outer boundary conditions corresponding to an (i) ideal spin sink, (ii) zero spin current, and (iii) a superlattice stack. Here  $d_{F}/\lambda_{m}=1$ and $d_{N}/\lambda_{N}=0.1$ in (a), and $d_{N}/\lambda_{N}=0.1$ and $\kappa_{F}\lambda_{N}/(\kappa_{N}\lambda_{m})=0.01$ in (b). (c) and (d): mMTR dependence on FM and NM thicknesses in a [FM$\vert$NM] superlattice with different NMs with $d_{F}/\lambda_{F}=0.1$ in (c) and $d_{N}/\lambda_{N}=0.1$ in (d). }
\label{Fig-2}%
\end{figure}

 \begin{table}
\caption{\label{tab:table1}
The parameters for several normal metal spacers with long spin diffusion lengths at room temperature.}
\begin{ruledtabular}
\begin{tabular}{cccccccc}
NM &$\sigma_{N}$(S/$\mu$m) & $\kappa_{N}$(W/(K$\cdot$m)) &$\lambda_{N}$(nm)&$G_{N}$($10^{16}\,$m$^{-2}$) \\
\hline
Al & $31$ &237& 600 \cite{PhysRevB.67.085319}  & 5 \\
Cu  & $35$ &401& 350 \cite{jedema2001electrical} & 10 \\
Au  & $19$ &318& 60 \cite{PhysRevB.72.014461} & 32
\end{tabular}
\end{ruledtabular}
\end{table}
We measure the heat-valve performance by the mMTR ratio $ (\rho^{\text{AP}}-\rho^{\text{P}})/\rho^{\text{P}}$ 
\begin{align}
\text{mMTR}&=\frac{\eta}{\frac{2d_{F}}{\lambda_{m}}+\frac{\kappa_{F}}{\kappa_{N}}\frac{d_{N}}{\lambda_{m}}+ 2\eta\frac{1+\xi\tanh\frac{d_{F}}{\lambda_{m}}\tanh\frac{d_{N}}{2\lambda_{N}}}{\coth\frac{d_{F}}{\lambda_{m}}+2\xi\coth\frac{d_{N}}{\lambda_{N}}+\xi^2\tanh\frac{d_{F}}{\lambda_{m}}} }\nonumber\\
&\times\frac{4\xi\tanh^2\frac{d_{F}}{\lambda_{m}}\text{csch}\frac{d_{N}}{\lambda_{N}}}{1+2\xi\tanh\frac{d_{F}}{\lambda_{m}}\coth\frac{d_{N}}{\lambda_{N}}+\xi^2\tanh^{2}\frac{d_{F}}{\lambda_{m}}},\label{mMTR1}
\end{align}
where $\xi=G_{N}/G_{m}$ is the ratio of the spin conductances of electrons to magnons and
\begin{align}
\eta=zT_{m} \frac{\kappa_{m}}{\kappa_{F}}\left(1-zT_{m} \frac{\kappa_{m}}{\kappa_{F}}\right)^{-1}
\end{align} 
is another dimensionless figure of merit that measures the impact of a magnon accumulation on the thermal conductivity. According to Eq.~(\ref{mMTR1}), $\eta$, $\xi$ and the normalized thicknesses of the NM and FM layers govern the magnitude of the mMTR that increases monotonically with increasing $\eta$ or decreasing $d_{N}/\lambda_{N}$, but decreases for large  $d_{F}/\lambda_{m}$ and $\xi$. In Fig.~\ref{Fig-2}(a) and (b) we tune the parameters to maximize the effect. When $d_{N}\ll\lambda_{N}$, the mMTR ratio approaches a $\xi$-independent maximum value
\begin{align}
   \text{mMTR} \overset{d_{N}\ll\lambda_{N}}{\rightarrow}\frac{\eta}{1+\frac{R_{N}}{R_{F}}}\frac{\lambda_{m}}{d_{F}}\tanh\frac{d_{F}}{\lambda_{m}}\label{max}
\end{align}
where $R_{N}=d_{N}/\kappa_{N}$ and $R_{F}=2d_{F}/\kappa_{F}$ are the thermal resistances in the absence of the magnon accumulation

\emph{(ii) Zero spin current.} When the spin valve is sandwiched by nonmagnetic insulators (e.g., MgO), the magnon spin currents at the outer interfaces of two FM layers vanish, leading to
\begin{align}
\text{mMTR}&=\frac{\eta}{\frac{2d_{F}}{\lambda_{m}}+\frac{\kappa_{F}}{\kappa_{N}}\frac{d_{N}}{\lambda_{m}}+2\eta\frac{2\tanh\frac{d_{F}}{2\lambda_{m}}+\xi\coth\frac{d_{N}}{2\lambda_{N}}}{1+\xi\coth\frac{d_{F}}{\lambda_{m}}\coth\frac{d_{N}}{2\lambda_{N}}} }\nonumber\\
\times&\frac{2\xi\text{csch}^{2}\frac{d_{N}}{2\lambda_{N}}\tanh^{2}\frac{d_{F}}{2\lambda_{m}}}{(\xi\coth\frac{d_{F}}{\lambda_{m}}+\coth\frac{d_{N}}{2\lambda_{N}})(1+\xi\coth\frac{d_{F}}{\lambda_{m}}\coth\frac{d_{N}}{2\lambda_{N}})}\nonumber\\
&\overset{d_{N}\ll\lambda_{N}}{\rightarrow}\frac{\eta\frac{\lambda_{m}}{d_{F}} \tanh^{2}\frac{d_{F}}{2\lambda_{m}} \tanh\frac{d_{F}}{\lambda_{m}}}{1+\frac{R_{N}}{R_{F}}+\eta\frac{\lambda_{m}}{d_{F}}\tanh\frac{d_{F}}{\lambda_{m}}},\label{mMTR2}
\end{align}
which agrees with Eq.~(\ref{mMTR1}) in the limit of $d_{F}/\lambda_{m}\gg 1$, as expected. However, an FM thickness comparable to $\lambda_{m}$ suppressed the effect [Fig.~\ref{Fig-2}(b)] because the magnon accumulation near the outer interfaces compensates for the ones near the FM$|$NM interfaces.

\emph{(iii) Superlattice condition.} In a spin valve embedded in a magnetic superlattice with [FM$|$NM] unit cell, the same surroundings for two interfaces of each FM layer correspond to outer boundary conditions when the magnon chemical potentials at the outer boundary are opposite to the ones at the FM$|$NM interfaces, i.e., $\mu_{1}(-d_{F})=-\mu_{1}(0)$ and $\mu_{2}(d_{N})=-\mu_{2}(d_{N}+d_{F})$.
   \begin{align}
&\text{mMTR}=\frac{\eta}{\frac{2d_{F}}{\lambda_{m}}+\frac{\kappa_{F}}{\kappa_{N}}\frac{d_{N}}{\lambda_{m}}+4\eta\frac{\tanh\frac{d_{F}}{2\lambda_{m}}}{1+\xi\coth\frac{d_{N}}{2\lambda_{N}}\tanh\frac{d_{F}}{2\lambda_{m}}} }\nonumber\\
&\frac{8\xi\tanh^2\frac{d_{F}}{2\lambda_{m}}\text{csch}\frac{d_{N}}{\lambda_{N}}}{(1+\xi\tanh\frac{d_{F}}{2\lambda_{m}}\tanh\frac{d_{N}}{2\lambda_{N}})(1+\xi\tanh\frac{d_{F}}{2\lambda_{m}}\coth\frac{d_{N}}{2\lambda_{N}})} \nonumber\\
&\overset{d_{N}\ll\lambda_{N}}{\rightarrow}\frac{2\eta}{1+\frac{R_{N}}{R_{F}}}\frac{\lambda_{m}}{d_{F}}\tanh\frac{d_{F}}{2\lambda_{m}}. \label{mMTR3}
\end{align} 
Since the magnon accumulation near both two interfaces of each FM layer contributes to the magnetothermal transport, the mMTR is enhanced compared to the previous cases [Fig.~\ref{Fig-2}(a) and (b)], in the limit of $d_{N}\ll \lambda_{N}$ and $d_{F}\gg\lambda_{m}$ the mMTR becomes two times larger than  Eq.~(\ref{max}). A similar enhancement of the spin Seebeck effect has been reported in metallic magnetic multilayers \cite{PhysRevB.92.220407}.

Let us consider a specific spin valve composed of NM (= Al, Cu, Au) and the magnetic insulator yttrium iron garnet (YIG) with well-known parameters at room temperature ($T_{0}=300\,$K): $\sigma_{m}=5\times10^5\,$S/m \cite{cornelissen2016magnon}, $\mathcal{L}_{m}/\sigma_{m}=110\, \mu$V/K \cite{NoteLm} and $\lambda_{m}=70\,$nm \cite{ Notespin}, $\kappa_{\text{YIG}}=6.6\,$W/(K$\cdot$m) \cite{mohmed2019magnetic}, which gives $\eta=0.38$ and $G_{m}=2.12\times 10^{16}\,$m$^{-2}$. Table \ref{tab:table1} summarizes the relevant parameters for Al, Cu, and Au. Fig.~\ref{Fig-2}(a) and (b) plots the mMTR ratio as a function of $\xi$ and $d_{F}/\lambda_{m}$ for the aforementioned outer boundary conditions, respectively. Fig.~\ref{Fig-2}(c) and (d) shows the mMTR with the superlattice boundary condition for Al, Cu, and Au spacers as a function of the FM and NM thicknesses, respectively. The mMTR decreases monotonically with the NM thickness; it is largest ($\sim\eta$) at an optimal thickness of the FM that is much smaller than its magnon diffusion length, approaching zero when $\kappa_F d_{N}/(\kappa_{N}\lambda_{m})\rightarrow 0$.

\emph{Conclusion.---}We predict a substantial room-temperature mMTR in spin valves with electrically insulating FM layers. We trace its origin to an interfacial temperature drop caused by a non-equilibrium magnon spin accumulation that depends on the relative orientation between two FM layers. The mMTR ratio can be engineered by the spin conductances, boundary conditions, and layer thicknesses. Our results may explain the giant MTR ratio in metallic multilayers \cite{nakayama2021above}. One might consider simply adding the electronic and magnonic 
contributions to the MTR. Indeed, magnons contribute to a large MTR but suppress the GMR \cite{PhysRevB.96.024449}, which might partly explain the observations. However, a thorough treatment requires better knowledge of the magnon transport parameters in ferromagnetic metals and the inclusion of the magnon-electron drag effect \cite{PhysRevLett.18.395,PhysRevB.13.2072,costache2012magnon,PhysRevB.94.144407,PhysRevB.96.024449}.  

\emph{Acknowledgment.---}The authors thank Y. Sakuraba, T. Hirai, F. Makino, and A. O. Leon for valuable discussions. P.T., K.U., and G.E.W.B. were supported by Grant-in-Aid for Scientific Research (S) (No. 22H04965) from JSPS KAKENHI, Japan. P. T. was also supported by Grant-in-Aid for Early-Career Scientists (No. 23K13050), K.U. by ERATO "Magnetic Thermal Management Materials Project" (No. JPMJER2201) from JST, Japan, and G.E.W.B. by JSPS KAKENHI Grants (Nos. 19H00645 and 24H02231).

\bibliography{reference}

\begin{thebibliography}{46}
\bibitem [{Not()}]{Notecon}%
  \BibitemOpen
  \href@noop {} {}\bibinfo {note} {{This assumption is justified when the bulk
  spin and magnon resistances are much larger than those of the interfaces. For
  the interface of YIG and typical normal metals (e.g., Au, Pt), the
  interfacial spin conductance governed by the spin mixing conductance is of
  the order $10^{18} $-$ 10^{19}\,\text{m}^{-2}$ \cite{PhysRevB.92.054437},
  which is around two orders of magnitude of larger than bulk spin
  conductances.}}\BibitemShut {Stop}%
\end{thebibliography}

\end{document}